# Assessing New Hires' Programming Productivity Through UMETRIX - An Industry Case Study

Sai Anirudh Karre, Neeraj Mathur, Y.Raghu Reddy
Software Engineering Research Centre, IIIT Hyderabad, India
saianirudh.karri@research.iiit.ac.in, neeraj.mathur@alumni.iiit.ac.in, raghu.reddy@iiit.ac.in

*Abstract*—New hires (novice or experienced) usually undergo an onboarding program for a specific period to get acquainted with the processes of the hiring organization to reach expected programming productivity levels. This paper presents a programming productivity framework developed as an outcome of a three-year-long industry study with small to medium-scale organizations using a usability evaluation and code recommendation tool, UMETRIX, to manage new hire programming productivity. We developed a programming productivity framework around this tool called "Utpada" Participating organizations expressed strong interest in relying on this programming productivity framework to assess the skill gap among new hires. It helped identify under-performers early and strategize their upskill plan per their business needs. The participating organizations have seen an 89% rise in quality code contributions by new hires during their probation period compared to traditional new hires'. This framework is reproducible for any new-hire team size and can be easily integrated into existing programming productivity improvement programs.

*Keywords-Software Productivity; New Hires; Industrial Practices; Software Developers*

## I. BACKGROUND

In 2018, we developed and patented an automated usability evaluation framework, **UMETRIX**, to detect code-level implementation of usability guidelines for mobile-based applications [5]. This approach uses source code analysis for automated usability assessment of mobile applications [5]. Fig. 1 illustrates the control flow of the UMETRIX [1]. UMETRIX accepts a mobile APK (android package kit) file and one or more validation case file(s) as input. The validation case file contains the code snippet linked with a usability guideline for detection. As a first step, the framework decompiles the apk file and prepares it for source code analysis. *'Validation Test Case Generator'* loads all the validation case files and links them with the *'Validation Execution Engine'* for source code analysis. Post execution, the framework provides a validation report on the count of correct implementation of usability guidelines via code analysis. In case of incorrect or no implementation of usability guidelines, the framework recommends code-snippets through *'Recommendation Engine'* to avoid usability issues. The *'Validation Case DB'* contains bundled set of validation cases written using an authoring tool and can be used for multiple usability code evaluations. The *'Metric DB'* stores the results of each validation execution, and the *'Code Snippet Bank'* stores the code snippets linked with respective usability guidelines for the recommendation. We evaluated the UMETRIX in 20 mobile app development companies stationed in India. We gathered improvement data on their usability indices [5] and assessed the impact of UMETRIX on mobile usability.

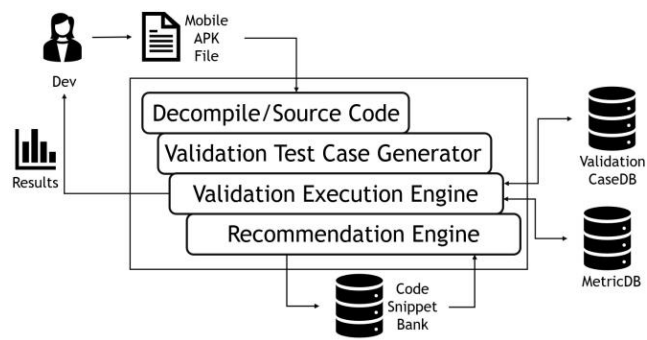

Figure 1. UMETRIX – Usability Evaluation Framework

> In contrast, these organizations leveraged UMETRIX as a **Programming Productivity Tool** to train and assess new-hire performance by expanding the **Code-Snippet Bank database**.

Following were the insights that led us to investigate this adaptation further.

- Organizations that employed UMETRIX saw a significant rise in the size of the Code Snippet Bank due to large set of code-snippet submissions over a short time.

- The code snippets were focused on the company-specific code-base. Thus these code snippets were unique for respective projects in a given organization.

- Development Teams started actively using the recommendation engine backed with Code Snippet Bank to train their new hires.

The above observations were pervasive across a few participating organizations, even though the product owners/managers addressed the issues differently. We ascertained this as an opportunity for UMETRIX tool [1] to be enhanced into a programming productivity framework to drive and track new-hire programming productivity. In this paper, we discuss the programming productivity framework called

"**Utpada**" backed by UMETRIX and its implementation across all the participating organizations following section.

### V. UTPADA - PROGRAMMING PRODUCTIVITY FRAMEWORK

**Programming Productivity** *is defined as the degree of the ability of individual programmers or development teams to build and evolve software systems* [3]. Over time, software organizations endeavor to improve their productivity factors by aligning them with overall product deliverables [9]. Product owners can anticipate tangible results in a given release cycle with enough FTE (full-time-equivalent) human resources by adopting different practices. Mentorship or Buddy programs are informal practices designed to train and nurture new hire programmers on coding standards, delivery policy, and SLAs defined in the given organization. The success of such programs will rely on mentor-mentee communication and interest in a common goal. The product owners may only guarantee better success from such programs if they depend on individual personalities. Considering these prevalent challenges, we discuss a case study of proposed "*Utpada*" - programming productivity framework backed by UMETRIX to train, upskill, and assess the productivity of new hire programmers. Fig 2. illustrates the "*Utpada*" - programming productivity framework that utilizes aspects of code snippet search and code snipped bank database that was part of the UMETRIX tool. Following are details about the necessary and sufficient conditions to execute this programming productivity framework.

- **Pre-Condition:** Current employers, including developers, UX practitioners, and QA practitioners, should be populating the Code-Snippet bank database of the UMETRIX tool with respective code-snippets widely used as part of their existing product code base.

- **New-Hires Usage:** When New Hires are onboarded, they are assigned new development tasks, including new feature stories and open defects. The new hires use Code-Snippet Search to use the best-fit code pattern for their deliverable. They re-use or re-code using the best fit code-snippet examples and submit the beta code for code review.

- **Review-Satisfaction-Index (RSI) Scores:** The mentors oversee the code review process and approve the beta code for check-in to production code if all deliverable requirements are met. If the beta code doesn't meet the requirements, they push it back to the new hire for re-work. The mentors submit a Review-Satisfaction-Index (RSI) [2], a code-review scorecard that captures the code snippet bank usage and other organization code standards ratings during code-review.

- **Post-Conditions:** The mentor, who acts as a code reviewer, submits the unique code snippets that new hires write for future curation into the code snippet bank database.

We developed an RSI score template to ease the code-review process in the context of this framework. The approved code check-in rate, RSI scores, and other programming productivity metrics are captured during the code review. These metrics help the respective Program Managers to track new-hire productivity. We approached all the organizations that have previously participated in our UMETRIX study to use the programming productivity framework. Some of these organizations participated in our study to review new hire programming productivity of both fresh graduates and experienced new hires. The details of the empirical study is provided in the following sub-sections.

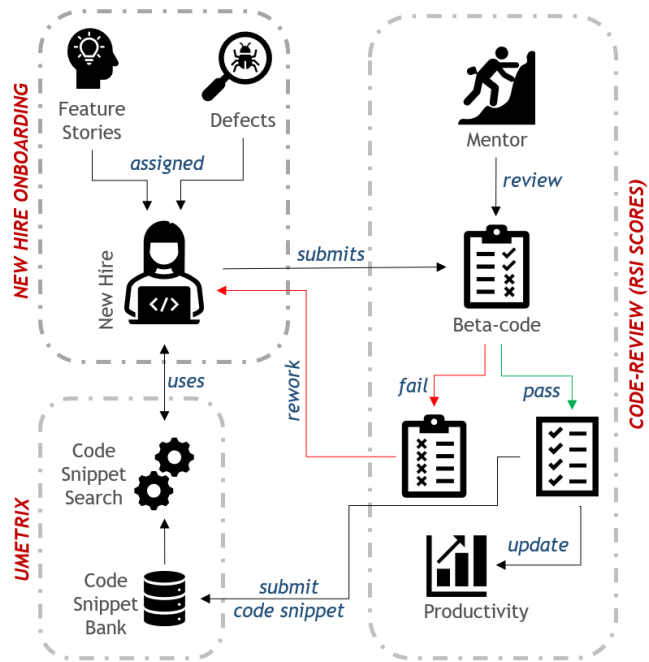

Figure 2. "**Utpada**" - New-Hire Programming Productivity Framework

#### A. Study Design

This section presents details about the participating organizations, demography of the new hires, study duration, programming productivity metrics, data tracking, and probation evaluation criteria, along with feedback from the participating organizations.

**Methodology:** A new hire had to undergo three steps to complete the onboarding and was deemed a "productive resource." *Step 1:* a new hire is mapped with a mentor who is an existing employee for training and onboarding purposes. *Step 2:* They are tasked with programming deliverables where the mentors conduct code reviews and capture RSI Scores. *Step 3*: the mentor and the following technical lead will review the RSI scores for about six months (sprint by sprint basis; 1 sprint = 6 working days) to move the new hire onto production.

**Data Collection:** We received programming productivity data every quarter. However, there was a delay in obtaining consent from the organizations towards the publication of programming productivity data as they come under Employee Monitory Laws of the United States, India, and Australia. As per US Federal Law, employers have the right to monitor their employees as they perform their duties. However, *US Electronic Communications Privacy Act of 1986 (ECPA)}, the US Common Law Protections Against the Invasion of Privacy,*

TABLE 1. DETAILS OF PARTICIPATING ORGANIZATIONS

| [Details][Company] | C1 | C2 | C3 | C4 | C5 | C6 |
|---|---|---|---|---|---|---|
| **Company Size** | 1000-2000+ (Medium) | 5000-10000+ (Large) | 5000-10000+ (Large) | 100-500+ (Small) | 500-1000+ (Small) | 10-100+(Small) |
| **Office/Team Type** | Distributed, open office | Multi-national | Multi-national | Distributed, closed office | Multi-national | Closed office |
| **Developer Tools** | Fixed tools | Uniform tools | Uniform tools | Uniform tools | Fixed tools | Similar tools |
| **Development Paradigm** | Scrum, Kanban | Scaled Agile, Kanban, Extreme | Agile, Kanban, Lean | Pair, Kanban | Pair, Agile | Pair |
| **Product Offering** | Mostly Mobile, Web | Mostly Web, Mobile | Mobile, Web, Embedded | Embedded, Mobile | Mobile, Web | Mobile, VR, AR |
| **Code Storage** | Single source, monolithic | Multi-source, separate | Multi-source, separate | Multi-source, separate | Single source | Monolithic |
| **HQ Country** | USA | USA | Australia | India | India | India |
| **Participating Teams** | 3 Teams | 6 Teams | 4 Teams | 1 Team | 1 Team | 2 Teams |
| **Industry Type** | Services | Accounting, Finance, Defence | Engineering | Services | Healthcare | Services |

*Indian IT Act 2000, and section and section 43A of IT Amendment Act, 2008* restrict practitioners to publish the information about the organization's trademark unless we have a No-Objection Certificate from the employer and employees who participated in the study. As the programming productivity data contains PII (Personal Identifiable Information) data, all the organizations prohibited us from declaring the participants' demographics. In lieu of this, all the references to the trademark names and employee information in this case study are masked. The mentors and code-reviewers (sometimes both) capture the productivity metrics during the Code-review session. Different in-house/Open-source tools were used to capture code quality metrics and other productivity metrics. Reviews were graded based on the code-review checklist and RSI scores were provided for the participant. The participants usually spend 4 to 6 months under the probation period before moving onto production. This transition will be done based on their performance during their tenure in the productivity framework of these RSI score. If the RSI scores are higher, the mentors recommend the Group managers to on-board the participants onto production teams early. The participants with exceptional performance usually spent 3-4 months in probation period. Participants with average performance spend around six months in probation. The under-performing participants were either terminated or supported with different roles which have non-development activities like QA, DevOps, etc.

**Participating Organizations:** We reached out to all the 22 Industry partners who successfully used the UMETRIX tool in practice to use the new-hire programming productivity framework. Only 6 of the organizations agreed to participate in the programming productivity study. Table 1 provides organization details and the participating team size. A multi-year-controlled case study was initiated within the participating organizations to understand the usage and impact of the Code-Snippet Bank database to enhance new hires' programming productivity. We started the study in March 2018 and ended our data collection by August 2021. We received programming productivity information from these six organizations on a Quarter-to-Quarter basis. The respective Group Managers (Head of Engineering) agreed to take up the responsibility to collate and share the data in a specified format [2]. We met with the respective owners every six months to introspect and review the strategy of skilling new hires.

**Participating Teams and New-Hires':** New hires are new employees joined in a particular organization after a thorough hiring process. In our context, new hires' are both fresh graduates and experienced with prior employment from previous organizations. These new hires are usually organized into teams mentored by a senior employee with expertise in existing work, usually called mentors. The participating teams were medium-sized, with 20 developers per team. Technical Leads act as mentors/code reviewers. They report to Group Managers within Engineering Teams. **80%** of our study participants were fresh university graduates with no prior industry experience, and the rest were experienced programmers. Overall, **21%** of them were female, **69%** were male, and the rest did not identify themselves by gender. They were hired during quarterly hire cycles between February 2018 and January 2021. The new hires participated in their respective organization's boarding process. They were later provided access to alpha/beta code branches, required developer tools, knowledge base articles, relevant product code-base training, and instructor-led training on overall deliverables. These new hires were expected to work on programming languages like Objective C, C#, Swift, ReactJS, AngularJS, Rust, Lua, Kotlin, and other JS frameworks. Although the new hires are hired across different hire cycles irrespective of their experience levels, they are still exposed to similar onboarding processes; the productivity data is captured per the prescribed programming productivity indicators for review.

**Programming Productivity Indicators:** Large teams among the participating organizations followed unconventional customized programming productivity indicators in line with their departmental goals. However, all six organizations agreed to measure a few common programming productivity indicators to track the new hire's performance during this study. These indicators include Deliverable throughput (DT), Lines Changed (LC), Weighted Average Class Complexity Metric, Code Quality metrics (Nested Block Depth, Leaky Encapsulation, Weighted methods per class, Type Checking, Feature Envy, Dead Code, etc.), and some minor Agile metrics include - lead time, cycle time, and velocity. They are well-established programming productivity metrics traditionally captured using different code-editor tools. Apart from the programming productivity metrics, a Review Satisfaction Index (RSI) was designed to aid the participants in tracking the health of their programming productivity based on feedback from their peer code reviewers. The raw data of the evaluated metrics is made available as supplement data [2]. The RSI scores play a crucial role in monitoring the new hire's programming productivity. It is a score-based weighted average designated by the code-reviewer based on a code-review checklist [2]. **6.5** out of **10** was benchmarked as the minimum score for RSI. Anything below this benchmark was considered as not being productive. The mentors and designated code reviewers share the RSI scores weekly with the participants and discuss the areas for improvement. Under-performers were coached by the mentors so that they could improve over time.

Here is an example to illustrate how the new-hire participants work on code assignment and use the code-bank to address such requirements. A requirement with id **REQ-21890**, that requires to avoid content view based on resizing text on the mobile screen. The new-hire searched the initial code snippet with *extActAttributes* as a keyword search. Its relevant code changes are updated per the requirement in change *C-REQ-21890* to raise the width of I properties to **100%** and hide the overflow text.

```
1. <input    type="text"    class="form-control
   extActAttributes" id="aligncompetency"/>
2. .extActAttributes {
3. display: inline-block; width: 70%;
4. }
```

> **REQ-21890** - Certain users - in particular, users with low vision, cognitive disabilities, and motor impairments - often need to increase the size of content in order to more comfortably read or operate a web page. All modern desktop browsers offer users "full-page zoom", which increases the size of all page content, including text, graphics, and overall layout. CONTENT BECOMES OBSCURED OR DIFFICULT TO UNDERSTAND WHEN RESIZED. When layouts don't appropriately respond to changes in zoom, content can be clipped and unreadable.
>
> In sample 04 - User Profile Page - Personal View - Create Task, the values of read-only fields in the "Attributes" section can be clipped if the text exceeds the width of the <input> elements due to layout not appropriately adjusting to allow for 100% width.

> **C-REQ-21890** Ensure that pages can be resized/zoomed to at least 200\% without any loss of content or functionality. In general, avoid the use of CSS absolute or sticky positioning, container dimensions set using viewport units like vh or vw, and containers which truncate content using overflow: hidden.
>
> In sample 04 - User Profile Page - Personal View - Create Task, set the width of the extActAttributes class to "100\%" for the fields in the "Attributes" section.

```
1. <input    type="text"    class="form-control
   extActAttributes" id="aligncompetency"/>
2. .extActAttributes {
3. display: inline-block;
4. overflow: hidden;
5. width: 70%;
6. }
```

```
1. fun main() {
2. try {
3. throw     IllegalStateException("Incorrect
   Typecast")
4. println("State Exception: Check REQ-1289")
5. }
6. catch (exc: Throwable){
7. println("DEBUG: Something went wrong")
8. }
9. }
```

**Code Review Proforma:** Michaela et al. extensively proposed an exhaustive code-review template for building proficient programmers [4]. This proforma covers aspects of code development like *Implementation* - with a focus on relevance and scope of the feature to be made, *Dependencies* - with a focus on compatibility and integrity of the feature, *Security* - with a focus on vulnerabilities, input sanitation, and validation, *Logic Errors* - events of code break and intention behind a code logic, *Error Handling* - logging and error management, *Usability/Accessibility* - with a focus of regulatory guidelines, *Performance* - with a focus on code run time and impact on overall system performance, *Readability* - with a focus on coding standards, code ethics, and code restructuring. Code Reviewer also captured the programming productivity metrics using their internal tools like Slack, GitHub, GanttPro, bitbucket, etc., from new hires of participating organizations. Below is a simple example of how to code reviewers expect error handling from new hires while building mobile features. The error handling refers to the exception to requirement **REQ-1289**. This helps the new hire programmer and the reviewer revisit the requirement and address the actual requirement.

**Data Collection:** The mentors and code-reviewers (sometimes both) capture the productivity metrics during the Code-review session. Different in-house/Open source tools were used to capture code quality metrics and other productivity metrics. Reviews were graded based on the code-review checklist and RSI scores were provided for the participant. The participants usually spend 4 to 6 months under the probation period before moving onto production. This transition will be done based on their performance of these RSI score. If the RSI scores are higher, the mentors recommend the Group managers to on-board the participants onto production teams early. The participants with exceptional performance usually spent 3-4 months in probation period. Participants with average performance spend around six months in probation. The under-performing participants were either terminated or supported with different roles which have non-development activities like QA, DevOps, etc.

*B. Results*

This section illustrates the results of the data points captured as part of this case study. Two types of data points are captured as part of this study. (1) RSI Scores through Code-review and (2) Code Bank Usage during code contribution.

- The Code-reviews generate RSI scores for the study participants on individual code contributions. The RSI combines a weighted score of programming productivity metric and a weighted score of code-review feedback, each of which is given *50* points. Overall the RSI is scored against *100* points and is normalized to a scale of 10 to ease the review.

- Code bank usage details are captured to understand the significance of code bank recommended code snippets while submitting code contributions.

- Code check-in details are captured by respective team leads using tools like *git, gitclear, and bitbucket*. Code Commit hits (counts), Line-Of-Code (LOC) added/updated/deleted, and Line Impact metric data daily as part of this study. In our study, no significant inferences could be drawn from this data. Hence it has not been included in the final review analysis.

*C. Discussion*

We present overall programming productivity observations as part of this section. The detailed programming productivity metrics and the code bank usage of all organizations are included as part of supplement data [2].

**Programming Productivity Conclusions**: Out of *101* new hire participants across all participating organizations, five new hires are terminated from the program due to their poor performance or lack of commitment to the training program. These programmers are employed with small participatory organizations. Thus all the code contributions of only *96* study participants are considered for this study. Among these participants, *77* exhibited exceptional performance in implementing the assigned deliverables. These study participants followed all the prescribed guidelines and excelled in their code contributions. Almost half received a perfect *10/10* RSI score for specific crucial code contributions during the code reviews. These exceptional new hires are transitioned to production-level code contributions based on recommendations from respective group managers. The average time these exceptional code contributors spend is around *14* weeks. Among the rest of the *19* study participants, *13* are moderate performers due to their focused knowledge of specific technologies. These participants displayed good code contributions in only distinct features and could not finish a few features they were not skilled enough. The average time spent by these medium code contributors is around *18-24* weeks. The rest of the *6* study participants are under-performers who could not deliver feature stories or defects during this case study on time. These participants are transitioned to non-functional teams like Datacenter, QA, and DevOps as they are technically equipped to manage non-functional aspects of the product. All of these under-performers are employed with large participatory organizations.

**UMETRIX CodeBank Usage Rate:** Overall *2002* code contributions are recorded by *96* study participants during the study period. Out of these code check-ins, *1921* code contributions relied on the code bank of UMETRIX to implement a feature story or a defect assigned. This data is captured as part of Code-review. While submitting the code contribution for code review, the new-hire participant must provide the CodeSnippetID of the code snippet utilized from the code bank. This code snippet is verified and used for reviewing the submitted code. The mentor or code review discusses the practical code-snippets relevance, implementation, and performance. This enables micro-learning for new-hire participants' during the code-review session. Code-reviewers observed that *89%* of the time, new-hire programmers relied on the right code-snippet for their code contributions i.e., *1710* code contributions out of *2002* code contributions across all the participating organizations directly relied on CodeBank recommendations. The detailed metrics are illustrated as part of supplement material [2].

**Impact on Software Quality:** Medium and Small participating organizations *C1, C4, C5* and *C6* have let the participants contribute to the product code-base after two weeks of new hire performance. The new hire developers have taken the mentor's feedback constructively and improved over time. The product owners observed *61%* rise in quality code contributions with only *4%* fatal errors in new hire code contributions. These organizations traced the overall quality metrics and evaluated them based on their internal departmental key performance indicator (KPI)s. Smaller organizations *C5* and *C6* with smaller teams correlated their current FY-19-20 quality metrics with FY2017-18 and observed a *33%* decrease in fatal errors and *55%* in trivial mistakes in production code. Large organizations *C2* and *C3* are reluctant about sharing the data on the improvement of overall product quality based on new hire code contribution. The detailed metrics are illustrated as part of supplement material [2].

**Replicating our Study** - Fig 2 illustrates the overall process of our study. The process is flexible enough to adapt and implement. The fundamental requirement of our study is to deploy the UMETRIX tool [1] and populate the Code Snippet bank database with relevant code snippets. The rest of the steps can be customizable. Tracking RSI scores and code-bank usage through new-hire code reviews will help engineering managers understand the programming productivity of new-hires.

*D. Perception Study*

We conducted an Onboarding Perception survey as part of a feedback session in October 2020. Twelve team leads and six group managers from the participatory organizations participated in the study. Fifty-two new hire programmers are also part of it. We conducted formal focused interviews with these voluntary participants to understand the impact and the consequence of our case study on these new hires and management during their tenure at these organizations. Detailed questionnaire can be found as part of our supplementary material [2]. Following are brief observations from the team leads and managers who played a managerial role in this study.

- With the RSI scoring method, code reviews helped new hires with frequent feedback. Code Snippet Banks are beneficial for persuasive knowledge transfer that includes in-house coding standards, design patterns, in-house metric benchmarks, and regulatory aspects of code during the early stage of product development.

- Few mentors used high RSI score-based code deliverables as a brown-bag training session to up-skill underperformers.

- Mentor-new hires pair programming sessions have improved work culture and created a beneficial synergy among the teams. Filtration of underperformers has also become easy and helped development managers identify submission risks early in code deliverables.

Following are the brief observations from participant new-hire programmers (both novice and experienced).

- Code Snippet bank found to be useful for new-hire developers to get accustomed to prevailing code faster than ever. The programmers also found the recommender engine of UMTERIX played an extraordinary impact on their RSI scores.

- All the new-hire programmers found UMETRIX easy to use. Experienced new-hire programmers found the process of code reviews to be elaborate and cumbersome to qualify.

- For most of them, this productivity framework provided clarity of thought on focusing on targeted deliverables. It helped address interruptions and distractions while working towards a deliverable.

### III. RELATED WORK

Developer programming productivity is one of the central aspects of First NATO's Software Engineering Conference (1968) [6]. Stefan et al. have illustrated the shift in proposals in their comprehensive review study about prevailing developer programming productivity Factors in Software Development [9] over decades. They could clearly define the distinction between the product, process and development environment to understand programming productivity consequences. Melanie et al. developed a ProdFLOW approach to capture automated programming productivity data from contributors of R&D organizations at Siemens AG [8] Chris et al. have also proposed unique metrics to track programmer productivity [7] and its implications on software quality. These two studies are widely distinctive as they illustrate different tool-based means to record and assess developer programming productivity. With vibrant studies on developer programming productivity, very few studies discuss the tool-based approaches to upskill, improve and track the programming productivity of new-hire programmers. Thus, we intend to present one such programming productivity framework **"Utpada"** using our existing UMETRIX tool to fill such a gap. We do not claim this is the only solution but suggest this as an alternate view towards assessing programmer productivity in software industries.

### IV. THREATS TO VALIDITY

This paper only illustrates a case study on how a usability evaluation tool (UMETRIX) was leveraged into a programming productivity tool (UTPADA) over time. Our study only presents the experiences of software practitioners who adopted and practiced a different approach to address the challenges involved with their new hire programming productivity. We have equally disclosed the scope of the UMETRIX tool to all the new-hire participants from the participatory organization. The Code snippet bank of the UMETRIX tool is exhaustive enough for participants to undergo the case study. The technical leads and group managers of the participatory organizations have taken enough care to update the code bank promptly to pick up the pace of code snippet usage. The mentor of the study participants stayed constant. Still, the role of the code reviewer rotates every week, avoiding bias in code-review data collection. Our study results are strictly captured under the supervision of technical leads at respective organizations. All our conclusions are based on the code review data captured by the mentors from the participatory organization.

### V. CONCLUSION AND FUTURE WORK

This paper discusses a new-hire programming productivity framework **"Utpada"** using the UMETRIX tool. The new hires use this framework as a self-help code snippet search to work on an enterprise source code base. **96** novice and experienced new-hire programmers participated in this multi-year case study. Code reviewers followed our programming productivity framework and captured RSI Scores to consider the productivity rates of new hires. Our study shows that **89%** of new-hire code contributions relied on the code snippet bank for quality code submissions to improve developer programming productivity. Events after the COVID-19 outbreak created some delays in data sharing. Thus we concluded the study by August 2021. As part of our future work, we planned to develop a code-visualization connector for UMETRIX as our future work for a better understanding of the control flow and data flow of code snippets to ease Code Bank usage.

ACKNOWLEDGMENTS

We thank all the new-hires' and participating organizations including their HR and Legal teams for executing and sharing programming productivity data as part of this study.